\documentclass[sigconf, authorversion]{acmart}

\copyrightyear{2024}
\acmYear{2024}
\setcopyright{acmlicensed}\acmConference[UIST '24]{The 37th Annual ACM Symposium on User Interface Software and Technology}{October 13--16, 2024}{Pittsburgh, PA, USA}
\acmBooktitle{The 37th Annual ACM Symposium on User Interface Software and Technology (UIST '24), October 13--16, 2024, Pittsburgh, PA, USA}
\acmDOI{10.1145/3654777.3676366}
\acmISBN{979-8-4007-0628-8/24/10}

\usepackage{xspace,array}

\newcommand{\system}{DiscipLink\xspace} 
\newcommand{\task}{IIS\xspace} 
\newcommand{\LLM}{LLMs\xspace} 
\newcommand{\search}{Semantic Scholar\xspace} 
\newcommand{\revision}[1]{{#1}}

\begin{document}

\title[DiscipLink]{DiscipLink: Unfolding Interdisciplinary Information Seeking Process via Human-AI Co-Exploration}

\author{Chengbo Zheng}
\email{cb.zheng@connect.ust.hk}
\orcid{0000-0003-0226-9399}
\affiliation{%
  \institution{Hong Kong University of Science and Technology}
  \city{Hong Kong}
  \country{China}
}

\author{Yuanhao Zhang}
\email{yzhangiy@connect.ust.hk}
\orcid{0000-0001-8263-1823}
\affiliation{%
  \institution{Hong Kong University of Science and Technology}
  \city{Hong Kong}
  \country{China}
}

\author{Zeyu Hunag}
\email{zhuangbi@connect.ust.hk}
\orcid{0000-0001-8199-071X}
\affiliation{%
  \institution{Hong Kong University of Science and Technology}
  \city{Hong Kong}
  \country{China}
}

\author{Chuhan Shi}
\email{chuhanshi@seu.edu.cn}
\orcid{0000-0002-3370-1626}
\affiliation{%
  \institution{Southeast University}
  \city{Nanjing}
  \country{China}
}

\author{Minrui Xu}
\email{mxubh@connect.ust.hk}
\orcid{0009-0001-0591-7526}
\affiliation{%
  \institution{Hong Kong University of Science and Technology}
  \city{Hong Kong}
  \country{China}
}

\author{Xiaojuan Ma}
\email{mxj@cse.ust.hk}
\orcid{0000-0002-9847-7784}
\affiliation{%
  \institution{Hong Kong University of Science and Technology}
  \city{Hong Kong}
  \country{China}
}
\renewcommand{\shortauthors}{Zheng, et al.}

\begin{abstract}
  Interdisciplinary studies often require researchers to explore literature in diverse branches of knowledge. 
  Yet, navigating through the highly scattered knowledge from unfamiliar disciplines poses a significant challenge. 
  In this paper, we introduce \system, a novel interactive system that facilitates collaboration between researchers and large language models (LLMs) in interdisciplinary information seeking (\task).
  Based on users' topic of interest, \system initiates exploratory questions from the perspectives of possible relevant fields of study, and users can further tailor these questions.
  \system then supports users in searching and screening papers under selected questions by automatically expanding queries with disciplinary-specific terminologies, extracting themes from retrieved papers, and highlighting the connections between papers and questions.
  Our evaluation, comprising a within-subject comparative experiment and an open-ended exploratory study, reveals that \system can effectively support researchers in breaking down disciplinary boundaries and integrating scattered knowledge in diverse fields. 
  The findings underscore the potential of LLM-powered tools in fostering information-seeking practices and bolstering interdisciplinary research.
\end{abstract}





\maketitle

\section{Introduction}


The knowledge of humankind is classified into disciplines, and many researchers are trained and specialize in a single discipline~\cite{abbott2010chaos}. 
However, there is a growing trend towards adopting interdisciplinary approaches to research, especially for complex, real-world problems~\cite{van2015interdisciplinary}. 
Although forming research teams composed of experts from various fields is one preferred way of carrying out interdisciplinary studies, resource constraints and budget limitations sometimes make it impractical to establish such teams~\cite{brown2015interdisciplinarity}. 
Researchers trained in one discipline often have to independently seek out information from the literature from various, maybe unfamiliar, research fields to develop their project ideas and approaches, a process we refer to as interdisciplinary information seeking (\task) in this paper.
However, this process is challenging for several reasons.
First, relevant literature to an interdisciplinary project is often \textbf{highly scattered} in various weak-connected research areas~\cite{mote1962reasons, bates1996learning}.
It often requires multiple iterations for researchers to identify \textit{what to search}~\cite{foster2004nonlinear, palmer2010information}.
Second, different disciplines use \textbf{different terminologies}, which becomes an obstacle for researchers to form accurate queries for searching literature in their unfamiliar fields~\cite{foster2004nonlinear, newby2011entering, spanner2001border}.
Also, interdisciplinary researchers face the challenge of interpreting \textbf{vast amounts of information} from various unfamiliar fields and \textbf{evaluating the relevance} to their interests during the search~\cite{foster2004nonlinear, palmer2010information}.

\begin{figure*}
    \centering
    \includegraphics[width=0.95\linewidth]{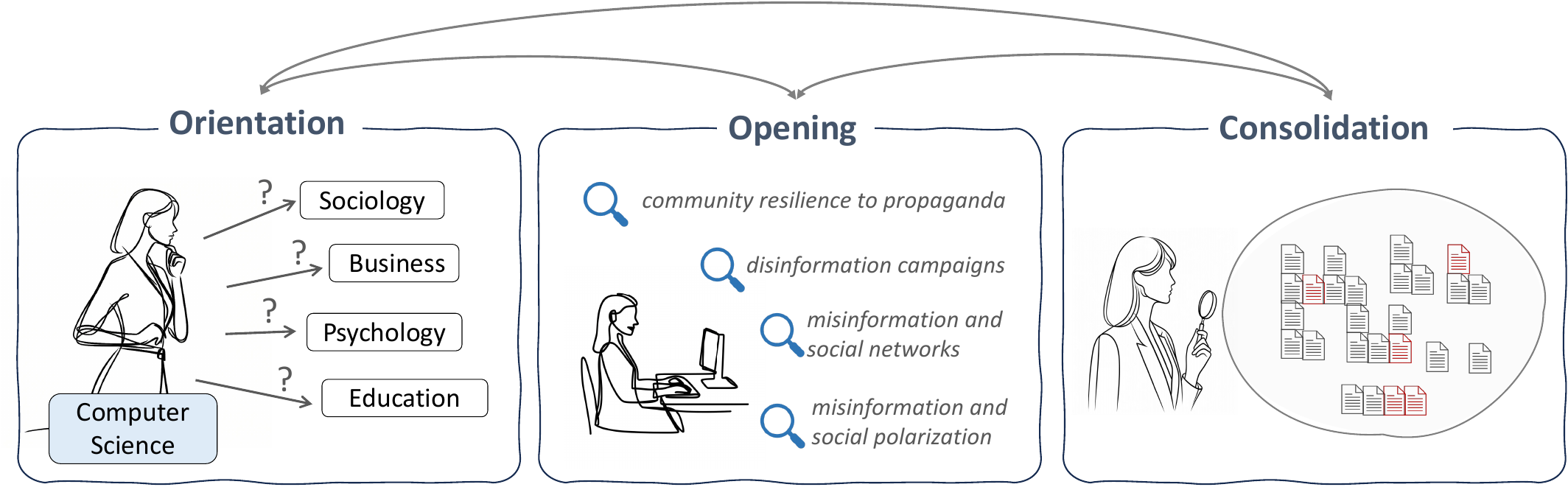}
    \caption{The nonlinear model of \task behavior includes three components~\cite{foster2004nonlinear}. Researchers do not go through these components with a certain order, but intertwine the activities in these components during \task. }
    \label{fig:nonlinear}
\end{figure*}

While prior human-computer interaction (HCI) work has proposed various approaches for seeking scholarly information, few of them are targeted at the needs of \task.
For example, when a computer science researcher is looking to find existing work on computational approaches to combat misinformation, existing literature search tools can facilitate the search by expanding on citations and references~\cite{chau2011apolo, kang2022threddy, kang2023synergi}, iterating search keywords~\cite{choe2021papers101}, or tracking key actor activities~\cite{kang2023comlittee}.
However, when the researcher is curious about what knowledge in sociology and education, in particular, can be blended into research on this topic, these tools provide limited guidance within the vast search space.
In this paper, we aim to leverage the recent breakthroughs in Large Language Models (LLMs) to tackle the challenges of \task. 
LLMs, such as GPT4~\footnote{\url{https://openai.com/gpt-4}} and Gemini~\footnote{\url{https://deepmind.google/technologies/gemini}}, are pre-trained on vast amounts of internet-wide text data, enabling them to embed the knowledge from various academic fields and can potentially tackle challenges mentioned above~\cite{brown2020language, liang2022holistic}.
Various LLM-powered literature search tools have emerged, such as Consensus~\footnote{\url{https://consensus.app/search/}}, Perplexity~\footnote{\url{https://www.perplexity.ai/}}, and SciSpace~\footnote{\url{https://scispace.com/}}.
These tools generally retrieve relevant papers based on user questions and provide answers grounded on the retrieved papers, which lowers the risks of hallucination~\cite{lewis2020retrieval}.
However, current LLM tools might still not be a perfect solution for \task for several reasons.
First, interdisciplinary projects often target ``ill-structured problem space''~\cite{palmer2010information}. 
As a result, the search process is by nature exploratory, and the intents of researchers will change along with the search process~\cite{palmer2010information}.
Automating the search with LLMs does not account for the shifts and leaves users with the burden of iterating exploration.
Second, existing workflows allocate the paper screening task entirely to LLMs.
However, the paper screening process is critical as interdisciplinary researchers need to make sense of outside domains~\cite{palmer2010information} continuously, have serendipitous discovery~\cite{foster2004nonlinear}, and build cognitive contexts~\cite{foster2004nonlinear, teevan2004perfect} from the process. 
Moreover, it is found that the information seeking by LLMs might be biased, creating a ``generative echo chamber''~\cite{sharma2024generative}.
Thus, it is critical to engage researchers in the exploration process instead of solely relying on the responses provided by LLMs.

To better assist researchers in \task with LLM, we design and develop \system, a human-centered AI system to \textit{co-explore} literature in various disciplines with users.
\system integrates a mixed-initiative~\cite{horvitz1999principles} and human-AI collaboration~\cite{zheng2022telling, shi2023retrolens} workflow.
With users' topic of interest as input, \system first generates \textit{exploratory questions} (EQs) around the topic from the perspectives of various relevant disciplines.
In this process, LLMs simulate domain experts from specific fields to provide EQs.
Users can tailor the EQs or elicit \system to generate more EQs by either making direct requests or sharing papers they find interesting, thus guiding the system to uncover new directions for exploration.
Besides, when users want to delve into any specific EQs, \system surfaces the existing knowledge about the EQs for users.
Specifically, we design an LLM-based query formulation strategy by including discipline-specific terminologies in the queries to extensively search papers around the EQs while maintaining relevance to the intended field(s).
After retrieving papers from queries, \system extracts key themes that are related to users' topics of interest as well as selected EQs to facilitate users in finding answers for the EQs.
Additionally, \system annotates returned papers by highlighting their connection with users' exploration focus to facilitate users in efficiently screening papers.

We evaluated \system through two complementary studies.
First, we tested the usability of \system with 12 graduate students by comparing our system with a baseline condition on searching papers for specific interdisciplinary research topics.
The quantitative findings indicate that \system enabled participants to complete their tasks more efficiently and to uncover a more comprehensive range of knowledge.
In the second study, we invited seven experienced interdisciplinary researchers to use \system for literature exploration in their own projects.
The qualitative feedback from participants highlighted their appreciation for \system's co-exploration workflow. 
Participants also identified potential limitations of \system, particularly in meeting the unique needs of individual researchers.

In summary, the contribution of this paper is threefold:

\begin{itemize}
    \item \system, an interactive system featuring a human-AI co-exploration workflow, empowering researchers in interdisciplinary information seeking.
    \item A comparative experiment and an open-ended exploratory study to evaluate the effectiveness of \system.
    \item Design implications derived from our design process and the user studies on how to support interdisciplinary researchers with LLMs.
\end{itemize}


\section{Related Work}


In this section, we first introduce the key processes and challenges of interdisciplinary information seeking.
Then, we discuss the existing systems that support the literature review.
Lastly, we discuss existing human-AI collaboration research on information-seeking and sensemaking.

\subsection{Interdisciplinary Information Seeking (\task)}
\label{rw:idsearch}

Interdisciplinary research involves integrating knowledge from multiple disciplines to address research problems~\cite{choi2006multidisciplinarity, van2015interdisciplinary}.
A critical component of this type of research is interdisciplinary information seeking (\task), the process of gathering relevant information across various fields.
Compared with normal scholarly information seeking, \task is typically more complex and time-consuming due to its highly dynamic process~\cite{spanner2001border, foster2004nonlinear}.

\citet{foster2004nonlinear} describe \task through a nonlinear model comprising three main stages: \textit{orientation}, \textit{opening}, and \textit{consolidation}, which are often intertwined during the \task process (see \autoref{fig:nonlinear}). 
\textit{Orientation} involves researchers determining their search direction.
The key challenge in \textit{orientation} is \textbf{the highly-scattered nature of the necessary knowledge for an interdisciplinary project}~\cite{foster2004nonlinear, newby2011entering, palmer2010information, bates1996learning}.
Researchers aim to construct an interdisciplinary knowledge picture for their projects, yet they cannot be experts in every field. 
Consequently, they often seek out resources, including experts, to discern relevant knowledge~\cite{knapp2012plugging}. 
In the absence of such guidance, they have to rely on their own general knowledge and intuition to determine the disciplines from which to gather information~\cite{foster2004nonlinear, palmer2010information}. 
This approach, however, introduces considerable uncertainty into their search process.
Furthermore, researchers often need to revisit their search goals based on insights from the \textit{opening} and \textit{consolidation} stages, adding to the time required for \task~\cite{foster2004nonlinear}.

\textit{Opening} is the stage where researchers explore literature in their chosen search directions.
Interdisciplinary researchers tend to \textit{deliberately increase the breadth of their search} and maintain an open-minded attitude towards all encountered information, considering it potentially valuable~\cite{foster2004nonlinear}. 
This approach is driven by the necessity to gather widely dispersed knowledge inherent to \task.
A significant challenge in this stage is the \textbf{terminology gap}: researchers often struggle to craft queries that use appropriate terminologies to effectively bridge their research interests with literature from unfamiliar disciplines~\cite{palmer2010information, bates1996learning}.
Previous work suggests that terminology may ``have a different meaning''~\cite{newby2011entering}, and ``not always appropriate or transferable''~\cite{foster2004nonlinear} across disciplines.
Researchers might not be able to immediately find the papers for the search direction they identified in \textit{orientation} due to the terminology gap.
Thus, they typically need to gradually accumulate their knowledge in unfamiliar fields and iterative refining search terms to uncover relevant literature~\cite{spanner2001border}.

\textit{Consolidation} is the stage where researchers decide whether and how to include found papers in their projects, typically following the \textit{Opening} and \textit{Orientation} stages. 
Two primary challenges emerge during consolidation. First, the \textit{Opening} stage results in the accumulation of a large volume of information, much of which may be irrelevant noise, thereby imposing a significant \textbf{information load} on researchers~\cite{palmer2010information, newby2011entering}. 
Second, the found papers have their own contexts in their belonging disciplines, and to sift, verify, and incorporate the found papers, the interdisciplinary researcher must \textbf{make sense of the connection between the found papers and their own project}
This task is challenging for researchers who are not accustomed to reading literature from unfamiliar fields~\cite{palmer2010information, spanner2001border}.

In summary, \task is challenging because of the highly scattered relevant knowledge, terminology gap, information load, and the difficulty in identifying relevant papers.
These challenges identified by prior work motivate our system design (see \autoref{sec:goals}).

\subsection{Interactive Systems Supporting Literature Review}
Getting a comprehensive understanding of prior works is generally challenging for researchers~\cite{bruce1994research, rowland2002overcoming}.
Although survey papers synthesizing previous works into themes are commonly used by researchers to grasp fields~\cite{jesson2011doing}, it is uncertain whether existing surveys can effectively cover all fields and capture rapidly emerging research topics. 
To overcome the obstacles associated with literature reviews, some studies have aimed to simplify the process of digesting research papers~\cite{head2021augmenting, chang2023citesee}, while others have focused on aiding researchers in compiling relevant literature through mechanisms like citation networks~\cite{kang2022you, tang2008arnetminer, kang2022threddy} and identification of key authors~\cite{kang2023comlittee}.

Specifically, \citet{kang2023synergi} outline two primary strategies employed by prior interactive tools to assist researchers in synthesizing literature reviews from extensive bodies of work: the \textit{bottom-up approach}, which supports researchers in developing key themes through improved navigation at the paper level, as seen in tools like Threddy~\cite{kang2022threddy} and Apolo~\cite{chau2011apolo}, and the \textit{top-down approach}, which offers an overarching view of the literature, exemplified by Metro Maps~\cite{shahaf2012metro}. 
Building on these concepts, \citet{kang2023synergi} propose for a mixed-initiative approach that combines clustering papers via citation networks to provide an overview for researchers while also allowing them to customize this overview from the bottom up. 
Inspired by \citet{kang2023synergi}, our system incorporates a mixed-initiative workflow once a collection of relevant papers has been assembled.
But differently, our design focuses on a more expansive search process across various disciplines to meet the unique demands of \task. Therefore, the system we propose takes users’ descriptions of their research interests as input, instead of using seed papers as in \citet{kang2023synergi}’s work.
Moreover, we specifically tailor the discovery and synthesis workflow to preserve the link between discovered papers and the user's contexts, considering the potential diverge between them in \task.



Previous studies have also endeavored to enable researchers to uncover relevant knowledge from other disciplines, such as developing analogical literature search engines that facilitate the discovery of papers from distant fields sharing similar research aims~\cite{chan2018solvent, kang2022augmenting}, directions or challenges~\cite{lahav2022search}.
Nevertheless, the analogical search has a clear goal in performing literature-based discovery~\cite{smalheiser2017rediscovering, swanson1996undiscovered, swanson1986undiscovered} to uncover solutions that could be transferred for solving the target research problem~\citet{kang2022augmenting}.
Differently, our work targets a more expansive search process (i.e., \task), and focuses on supporting researchers oriented to relevant research in various disciplines.
Additionally, \citet{shi2022medchemlens} introduce a visual analytics system designed to automatically collate and synthesize research data from several disciplines, aiding in the selection of research directions in experimental medicinal chemistry. While their visual analytics method is tailored to a specific interdisciplinary domain, our study explores the broader potential of LLMs to facilitate the general processes involved in \task.


\subsection{AI-powered Information Seeking and Sensemaking}

Numerous initiatives have been undertaken to develop AI systems aimed at enhancing information-seeking and sensemaking activities. 
For instance, \citet{qiu2022docflow} introduced DocFlow, a visual analytics system that facilitates document retrieval and categorization through natural language questions within the biomedical field. 
\citet{palani2022interweave} design in-context positioning of query suggestions and found it improves the exploratory search process.

The advent of large language models (LLMs) has positioned them as potential knowledge bases for information retrieval~\cite{alkhamissi2022review, yang2022empirical}. There is a growing body of research focused on refining the user experience in sourcing information from 
LLMs. For example, \citet{suh2023sensecape} argue that the typical linear conversational workflow found in many LLM-based tools falls short in aiding users to synthesize and comprehend information generated by LLMs. 
To address this, they advocate for the adoption of interactive hierarchical views to enrich information interaction practices. 
\citet{jiang2023graphologue} present a novel approach by organizing LLM responses into interactive graph-based diagrams, thereby streamlining the process for users to pinpoint the information they need.

While LLMs in the above works mainly serve by directly generating answers for information seekers, hallucination remains a significant issue. 
To address this, some research focuses on using LLMs to extract and summarize information from search results. 
Selenite, an LLM-powered system recently proposed by \citet{liu2023selenite}, employs LLMs to generate comprehensive overviews of options and criteria grounded in search results, guiding users through complex decisions.
Recent emerging works also see opportunities to leverage potential hallucinated LLM generations.
For example, \citet{gao2022precise} use LLMs to generate ``hypothetical documents'' for input queries to improve dense retrieval.
Similarly, \citet{son2024genquery} utilize LLM generation for visual search, but to help users better express their intents and refine their searches. 
\citet{lee2024paperweaver} connect retrieved papers with users’ existing collections using LLMs, which may include hallucinations, to aid in sensemaking. 
They found that in their design, hallucinations had minimal negative impacts.
In our work, we also explore how potentially hallucinated generations can support information seeking.
We use LLMs to produce high-level exploratory questions, instead of facts, from various disciplinary perspectives.
These questions enable users to co-explore with LLMs.
Additionally, we leverage LLMs to aggregate discipline-specific terminology by generating pseudo-answers, thereby expanding search queries with relevant terminologies to help users bridge terminology gaps in \task.






\section{Design Goals}
\label{sec:goals}

Drawing on insights from prior research on interdisciplinary information seeking (refer to \autoref{rw:idsearch}) and reflections from our design iterations (detailed in the supplementary material), we have identified four design goals (\textbf{DG}) essential for designing a human-AI collaborative system to aid in \task.

\begin{enumerate}
    \item[\textbf{DG1}] \textbf{Support orienting the exploration in various disciplines.}
    Engaging in interdisciplinary research often necessitates acquiring knowledge from various disciplines, which challenges researchers in determining relevant search areas in unfamiliar fields~\cite{palmer2010information, foster2004nonlinear, newby2011entering, bates1996learning}. 
    Although LLMs can assist researchers in delving into topics across various disciplines, their tendency to generate hallucinations may lead to confusion. 
    Additionally, prior studies~\cite{foster2004nonlinear, shah2022situating} and our preliminary design experiments underscore the importance of maintaining high self-agency for researchers who prefer to be actively involved in the exploration process. 
    Accordingly, the system should not solely provide answers for \task with LLMs but should utilize LLMs as a knowledge resource, guiding users in their exploration and enabling a collaborative exploration experience.

    \item[\textbf{DG2}] \textbf{Support formulating search queries with domain-specific terminologies.}
    Formulating effective search queries, especially with terminology spanning multiple fields, poses a significant challenge for researchers in \task~\cite{palmer2010information, newby2011entering}. 
    This often involves iterative refinement of search keywords, informed by ongoing discoveries from the literature~\cite{foster2004nonlinear}. Given that LLMs excel in linking concepts and generating diverse query sets~\cite{wang2023query2doc, wang2023can}, the system should leverage LLMs to assist users in crafting queries, thereby saving time and enhancing search efficiency.

    \item[\textbf{DG3}] \textbf{Organize retrieved papers around exploration focus.}
    The sheer volume of papers encountered in \task can overwhelm researchers~\cite{foster2004nonlinear, newby2011entering}. 
    Moreover, search results often include irrelevant noises that detract from the exploration focus, as revealed in our design iterations. 
    Previous HCI research suggests the aggregation of similar information and providing overviews is helpful in such cases to alleviate user burden~\cite{kang2023synergi,pirolli1996scatter, chau2011apolo}. 
    Therefore, the system should categorize retrieved papers into overviews, enabling users to swiftly pinpoint relevant groups of literature.

\item[\textbf{DG4}] \textbf{Provide Information Scent for Efficient Paper Sifting.}
In the \task process, researchers often need to meticulously review papers to verify their relevance to specific aspects of their research~\cite{palmer2010information, foster2004nonlinear}. 
This labor-intensive task necessitates a mechanism for quickly identifying relevant information within papers. 
To facilitate this process, the system should highlight critical paper details and offer ``information scents'' that guide researchers in their review process~\cite{pirolli1999information}.

\end{enumerate}



\begin{figure*}
    \centering
    \includegraphics[width=0.8\linewidth]{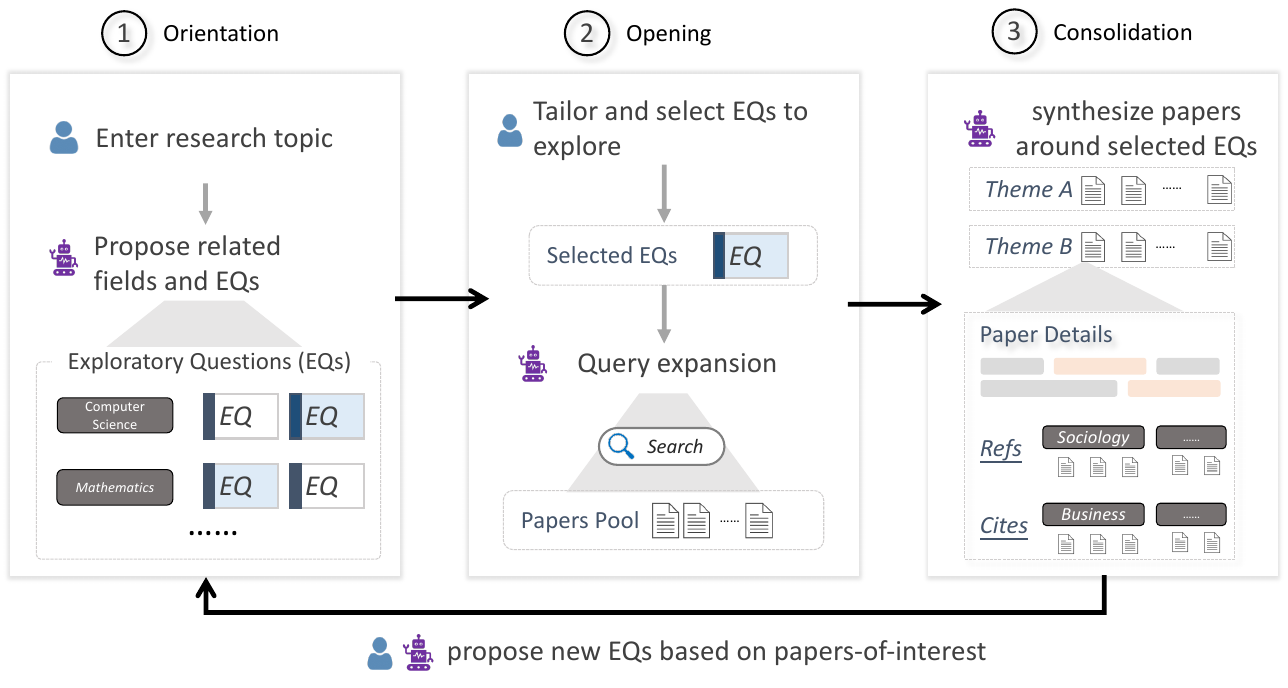}
    \caption{The workflow of \system. We harness LLMs to support the three stages of the \task process.}
    \label{fig:workflow}
\end{figure*}

\section{The \system System}
We introduce \system, a human-AI collaborative system designed to enhance interdisciplinary information seeking (\task). 
Unlike previous efforts in scholarly information seeking, \system focuses on enabling researchers to effectively collaborate with LLMs for broader exploration across disciplines, thereby linking their own projects with diverse knowledge.

Generally, inspired by prior research~\cite{kang2023synergi, horvitz1999principles}, \system facilitates both top-down and bottom-up mixed-initiative literature discovery workflows (\autoref{fig:workflow}).
In the top-down workflow, users decide the exploration direction they are interested in first, avoiding being lost in the details of a number of papers.
To aid this process, \system generates diverse EQs from various disciplines based on the user's research topic (\autoref{fig:workflow}-1), crafts comprehensive queries for each EQ (\autoref{fig:workflow}-2), and summarizes key themes from retrieved papers to streamline screening (\autoref{fig:workflow}-3).
On the other hand, the bottom-up approach empowers users to spontaneously navigate through papers, citations, and references.
Users can organize the papers of interest and formulate new EQs based on these papers, thereby starting new research directions.
\system enhances this exploration by organizing citations and references based on user exploration history and suggesting EQs informed by user-selected papers.

As a research prototype, \system uses GPT-4 as the underlying LLM, \search API service for searching papers, and ``\texttt{text-embedding-3-small}''~\citet{openai2024embedding} for text embedding.
The subsequent subsections detail a usage scenario and describe \system's key features in addressing our design goals.

\subsection{Usage Scenario}
Zoey, a computer science researcher, is tackling a new project aimed at raising awareness about social media misinformation among older adults. 
While she possesses expertise in misinformation detection through natural language processing, she finds herself stumped by the challenges older adults face in navigating social media and the human factors influencing misinformation awareness.
To bridge these gaps, Zoey turns her attention to unfamiliar disciplines like sociology and psychology for insights. 
Beginning with search terms such as ``older adults misinformation,'' through Google Scholar, she finds a mountain of papers that seem relevant across disciplines, including education, sociology, and psychology, each offering unique perspectives on the subject. 
Determined not to miss out on anything potentially useful, Zoey digs into each paper, one by one. 
However, the scattered insights across different fields make her research feel like assembling a vast puzzle without a guiding image.
Moreover, she needs to understand different perspectives during her search, each requiring a shift in mindset. 
Struggling to manage the information overload and the constant shifts in disciplinary perspectives, Zoey considers seeking help from \system.

Upon describing her research project to \system, Zoey receives a list of exploratory questions (EQs) organized by specific disciplines.
Zoey is particularly interested in the perspectives of psychology, education, and sociology.
These EQs, framed with easy-to-understand language, introduce some key factors, such as ``What can older adults be motivated to learn digital literacy?'' (education), ``What cognitive strategies reduce belief in false information?'' (psychology), that has never come to her mind before, possibly due to her unfamiliarity with these disciplines.
Among these, the EQ ``How effective are peer-led education programs for seniors?'' catches her interest. 
However, aiming to align it more closely with her focus on misinformation, she refines it to ``How can peer-led education programs combat misinformation among older adults.''

\begin{figure*}
    \centering
    \includegraphics[width=1\linewidth]{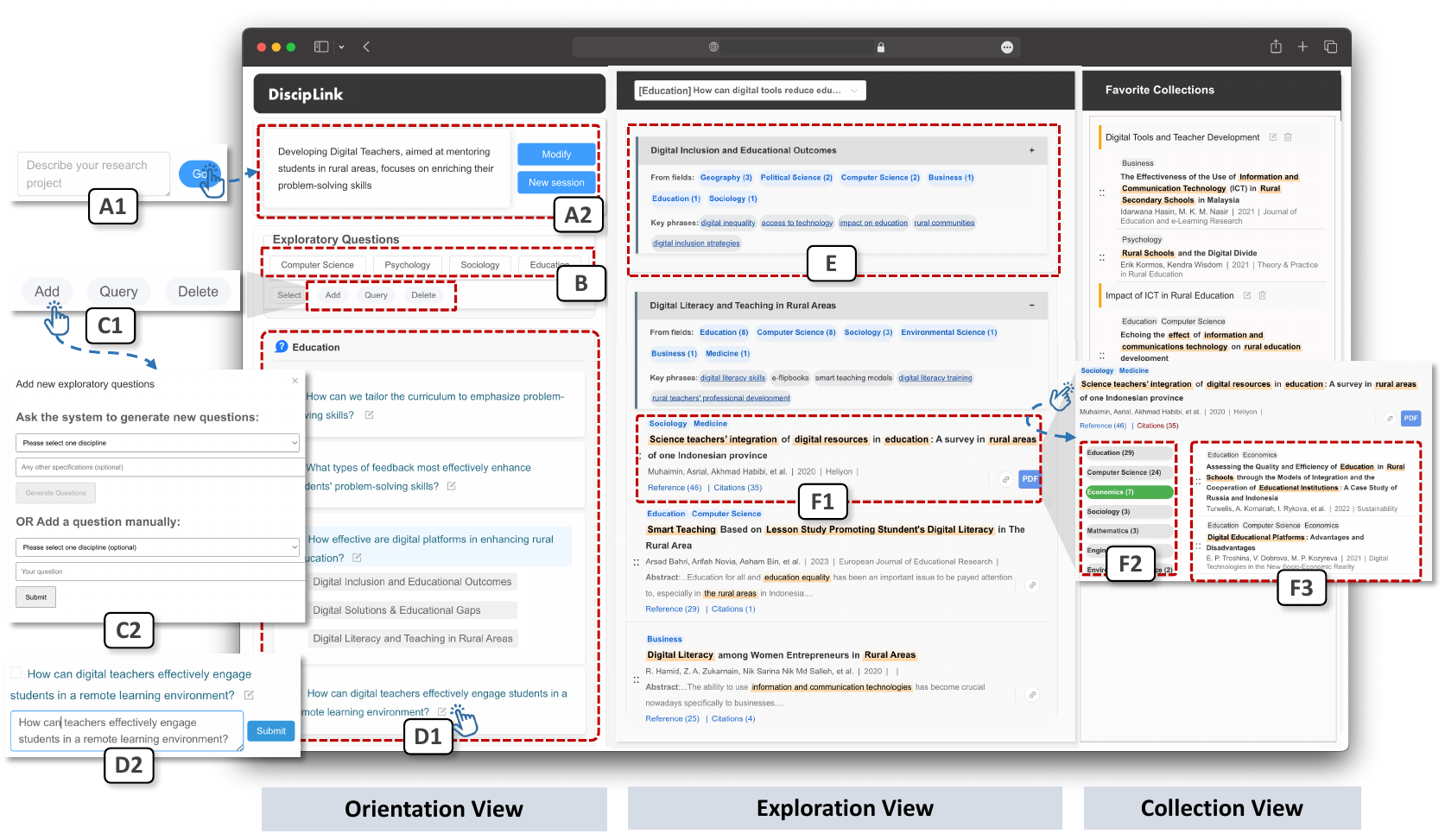}
    \caption{The user interface of \system, including \textit{Orientation View} for creating and tailoring EQs, \textit{Exploration View} for explore retrieved papers for each EQ, and \textit{Collection View} for organizing papers users found useful.}
    \label{fig:system}
\end{figure*}

After tailoring several EQs across psychology, education, and sociology to her preferences, Zoey prompts \system to fetch relevant papers. 
After searching, \system organizes the findings into themes for each selected EQ in the Exploration view (\autoref{fig:system}-middle).
Delving into the search results on cognitive strategies to counter misinformation, Zoey discovers themes such as ``Critical Thinking and Cognitive Improvement'' and ``Inoculation and Overconfidence Strategies to Combat Misinformation''.
This overview allows her to navigate the themes without getting bogged down in the details of individual papers.
Attracted by the concept of ``psychological inoculation,'', a term new to her, she explores a linked wiki page, uncovering its relevance in preemptively addressing misinformation. 
Through annotated key phrases in paper titles and abstracts, such as ``misperception'' and ``credibility,'' Zoey confirms the relevance of these papers, collecting them by dragging the theme to the Collection view (\autoref{fig:system}-right).

Later, Zoey's exploration leads her to a highly cited paper, and habitually, she checks its citations. 
\system categorizes these citations by discipline, with ``Medicine'' surprisingly ranking in the top three. 
Out of curiosity, she examines the citations under Medicine, and based on the highlighted terms, she finds a paper that discusses the impacts of health misinformation on older adults.
She realizes the strong connection between health misinformation and the older adults, but she hasn't explored this direction.
To extend her exploration, Zoey drags the paper to the orientation view (\autoref{fig:system}-left).
Based on the paper, \system recommends three new EQs around health misinformation's impact on older adults in the context of social media, thus broadening her exploration horizon.

\subsection{System Design}

\subsubsection{\textbf{[DG1] Elicit Exploration across Disciplines}}
To support users in expanding their exploration to relevant information scattered in various academic fields (\textit{DG1}), \system encapsulates potential literature search objectives into interactive elements named \textit{exploratory questions} (EQs).
In other words, \system proposes EQs from different disciplines' perspectives to elicit users' breath exploration.
For example, when exploring a research topic of facilitating older adults to learn about AI tools, possible EQs could be ``\textit{what cognitive challenges older adults may face in understanding complex concepts?}'', which concerns psychology, and ``\textit{how to design interactive tools to support learning for older adults}'', which is more related to HCI literature.
Below, we introduce the enabled interactions around EQs in \system, the design rationale, and the EQ-generation process.\\

\textbf{Supported Interactions around Exploratory Questions.}
Upon users entering a research topic (\autoref{fig:system}-A1), \system generates a list of EQs tailored to the \textit{research topic of interest}. 
In this process, we leverage the inherited knowledge of \LLM derived from its large-scale training data to ensure the coverage of diverse disciplines. 
Users have the flexibility to modify their topic description anytime, prompting \system to produce new EQs based on the updated description or to initiate a new exploration session by resetting the current ones (\autoref{fig:system}-A2). 
The EQs are categorized by their relevant disciplines (\autoref{fig:system}-D1), with a navigation bar allowing users to easily filter EQs by research field (\autoref{fig:system}-B). 
By selecting EQs, users can specify their exploration intentions for \system to identify literature closely aligned with their interests. 
Further interactions include editing EQs (\autoref{fig:system}-D2), adding or removing selected EQs (\autoref{fig:system}-C1), and generating additional EQs by specifying disciplines and particular needs or by manually entering questions (\autoref{fig:system}-C2).

Note that if users encounter an interesting paper on a yet-to-be-explored topic, they can obtain new EQs from \system based on the paper by dragging it from the exploration or collection view to the orientation view. 
Users have the option to pinpoint specific aspects of a paper, such as particular keywords, that pique their interest. 
Upon specifying these details, they can direct \system to suggest new EQs tailored to these focal points.
We set \system to provide three new EQs each time.
Users can choose one or several EQs to refine or opt to craft new EQs from scratch. 
Once these EQs are finalized and submitted, \system initiates a new round of search for papers concerning these EQs immediately.\\

\textbf{Design Rational.}
Our design of EQs draws inspiration from human-centered design principles, including the creation of a \textit{shared representation} to balance automation with user agency~\cite{heer2019agency}, and the \textit{reification of actions into objects} for direct manipulation~\cite{beaudouin2000reification}.
EQs act as a bridge between users and the system, offering an intuitive means for users to convey their interests and for \system to interpret and further translate these interests into complex queries, thereby fetching a broad spectrum of relevant papers. 
We have explored other forms of LLM assistance in the design process for guiding and expanding user exploration in \task.
Initially, our prototype leveraged \LLM to generate research topic statements for users. 
However, this approach led to two main issues: the topics often contained inaccuracies due to LLM hallucinations, and they served more as end products of exploration instead of inspirations for users to explore.  
Another alternative we considered was utilizing \LLM to generate simple queries, similar to those users might input into search engines like Google Scholar. This method, while straightforward, can be inefficient in \task.
Effective queries for \task necessitate the use of discipline-specific terminology, which could be challenging for users to understand. 
To address the problems in the above two methods, EQs in \system are framed as questions rather than facts, aiming to expand the exploration space, and users can flexibly customize EQs to \textit{co-explore} with LLMs. Also, EQs are crafted in user-friendly language as a high-level abstraction of research directions over low-level typical search queries. \\

\begin{figure}
    \centering
    \includegraphics[width=\linewidth]{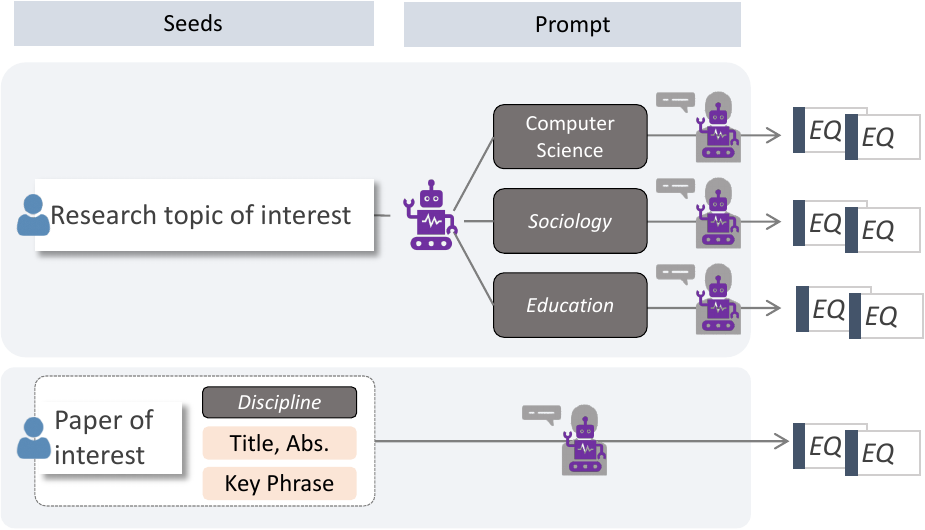}
    \caption{\system provides diverse EQ suggestions to discover highly-scattered relevant information based on two types of seeds given by users: research topic of interest and paper of interest. We prompt \LLM to simulate as experts in specific academic fields to provide EQs. Users can also create EQs by themselves.}
    \label{fig:eq-generation}
\end{figure}





\textbf{Prompts Design.}
\system takes two main types of inputs from users to generate EQs: the \textit{research topic of interest} and/or the \textit{paper of interest}, as illustrated in \autoref{fig:eq-generation}. To generate EQs from diverse disciplinary perspectives based on these seeds of exploration, we employ prompt-chaining and persona-prompting techniques~\cite{wu2022ai, white2023prompt}.

The process involves chaining prompts to guide \LLM in generating EQs. 
More specifically, given a research topic, we provide the list of disciplines used by \search for \LLM to determine which of them might be relevant.
The list includes primary disciplines, such as ``computer science'' and ``psychology''.
\LLM then determines specific subfields within these disciplines (e.g., ``developmental psychology'' within ``psychology'') that closely align with the topic of interest, enabling it to propose more targeted questions.
Then, for each identified field and subfield, our prompt to \LLM begins with, ``\textit{You are an expert in {field}. A researcher is developing a research topic: {topic}. What knowledge from your field could assist in developing this idea? Present your advice in the form of questions}.'' After collecting EQs across different disciplines, we instruct \LLM to refine this list by removing any duplicates, resulting in the final EQs shared with users.

Preliminary experiments indicate that adopting specific expert personas enables \LLM to produce questions that closely match the disciplinary focus.
For instance, when exploring how AI tools can support older adults  learning in science without employing a persona, language models tend to produce questions centered on ``\textit{AI for learning}.'' 
These questions focus predominantly on ``AI,'' steering the subsequent discovery process toward computer science literature rather than education. 
Conversely, utilizing persona prompting can generate EQs that lean more heavily towards aspects of older adults learning from an education perspective, such as ``\textit{What motivates older adult learners in science education?}''.

For generating EQs based on a \textit{paper of interest}, perhaps discovered through navigating the references and citations of papers returned in the previous rounds of exploration, \system adopts a similar persona-prompting strategy. 
Here, the meta-information of the paper, including its title, abstract, designated fields, and keywords, is integrated into the prompt context, allowing users to specify keywords that highlight their areas of interest. Following this, we prompt \LLM to identify relevant fields within the primary disciplines and simulate experts in these fields to recommend EQs related to the provided context. 

\subsubsection{\textbf{[DG2] Query Expansion to Retrieve Relevant Papers}}
\label{sec:query-expansion}
\begin{figure}
    \centering
\includegraphics[width=1\linewidth]{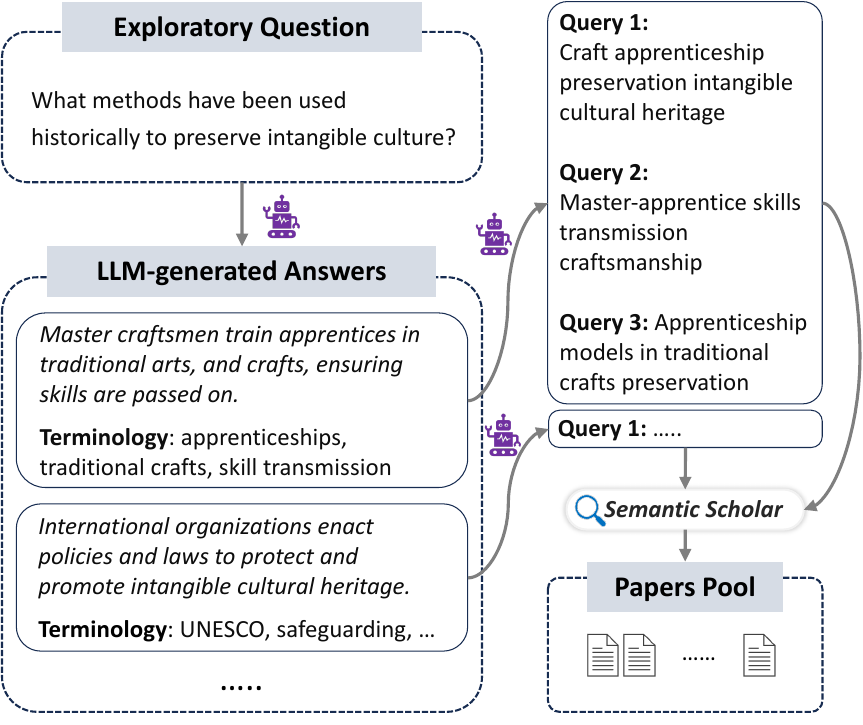}
    \caption{\system generates a diverse set of queries to retrieve relevant papers for an EQ.}
    \label{fig:query-expansion}
\end{figure}

When users express interest in a particular EQ, \system furthers the exploration by expansively generating numerous queries related to the EQ.
Next, it conducts searches for papers in \search using these queries concurrently.
In EQ generation, we guide \LLM to employ straightforward, easy-to-understand language so that users can quickly decide whether to dive into an EQ or not. 
However, when it comes to searching for papers, we instruct \LLM to use precise terminology and field-specific jargon relevant to the EQ to craft the search queries. 
This aims to fill the terminology gap of \task introduced in \autoref{rw:idsearch}.
Moreover, we intend to retrieve a diverse set of papers using the output queries to allow \system to map out the knowledge landscape around the EQ.
We design a prompt to craft queries according to these requirements, as shown in \autoref{fig:query-expansion}.

To be more specific, we begin by requesting \LLM to provide (pseudo-)answers from various perspectives on the current EQ in the form of bullet points. 
Although there is a risk of \LLM hallucination, these responses, even when flawed, are found to contain keywords and concepts beneficial for formulating effective search queries, as noted by \citet{wang2023query2doc}.
Subsequently, \LLM compiles a list of relevant terms associated with each bullet point. 
These terms lay the groundwork for creating a set of targeted search queries,  which are later fed into paper search engines.
This process, inspired by \citet{wang2023can}, helps to enrich the conceptual breadth and depth of inquiries into varied branches of knowledge.


\begin{figure*}
    \centering
    \includegraphics[width=1\linewidth]{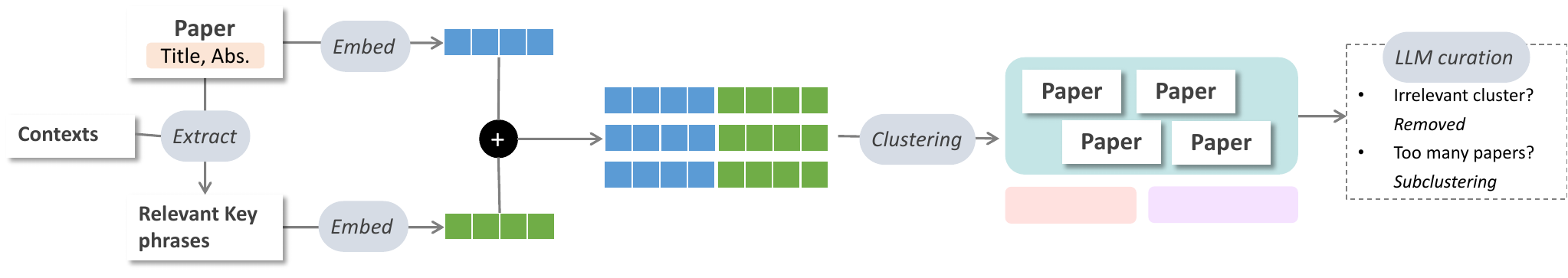}
    \caption{To organize the retrieved papers, \system first clusters the papers by embedding them based on the meta information and their connection to the exploration contexts (i.e., the research topic of interest and the associated EQ); then, \system uses \LLM to curate the clusters by removing less relevant clusters and doing subclustering if too many papers exist in one cluster.}
    \label{fig:cluster}
\end{figure*}

\subsubsection{\textbf{[DG3] Organize the Retrieved Papers of around EQs}}
\label{sec:organize}
It is challenging for users to go through the extensive papers retrieved from the queries, especially considering that many papers are likely from unfamiliar fields of study.
To alleviate this information overload, \system distills key themes from the papers, focusing on the connection to the user's current exploration context (i.e., the research topic and the selected exploratory question). \\ 

\textbf{Themes Extraction.} 
While the papers retrieved from the expansive queries could contain many themes, users would be more interested in those more related to their current exploration context, i.e., the research topic and the selected EQ.
A LLM-based clustering pipeline (\autoref{fig:cluster}) is deployed to sift through papers and extract themes pertinent to the user’s interests.
\revision{
Several LLM-based clustering techniques have been proposed by researchers, while trade-offs exist.
For example, LLooM~\cite{lam2024concept} and ClusterLLM~\cite{zhang2023clusterllm} produce high-quality clusters but need to operate at the data instance level -- LLooM involves distilling info from each input data instance and ClusterLLM uses triplet comparisons -- leading to increased time costs. 
To balance quality and time costs and enable interactive theme exploration in \system, our pipeline begins by generating contextualized embeddings for papers, proceeds to cluster these embeddings, and concludes with \LLM refining the clustering outcomes.
}

In the first step, we are inspired by the keyphrase expansion technique proposed by \citet{viswanathan2023large} to construct the contextualized paper embedding.
This technique concatenates original text embedding with keyphrase embedding related to user intents. 
Similarly, in \system, the contextualized embedding comprises text embedding of the paper's metadata (title and abstract) and embeddings of the paper's discipline, if matching the EQ's discipline, as well as key phrases from the paper's metadata resonating with the research topic and EQ. 
By considering whether the paper's discipline matches the EQ's discipline in the paper embedding, we add a soft constraint on clustering papers that fall into the same field of study. 
To identify relevant phrases from academic papers, we first convert key concepts \(C = \{c_1, c_2, ..., c_m\}\) from the research topic and EQs, along with phrases \(P = \{p_1, p_2, ..., p_n\}\) from the paper's metadata, into embeddings. 
A phrase \(p_i\) is considered relevant if its cosine similarity with any concept \(c_j\), defined as \(\text{sim}(v_{c_j}, v_{p_i}) > 0.6\), surpasses a predetermined threshold of 0.6~\footnote{This threshold is determined through empirical testing during development.}.

In the clustering phase, we employ DBSCAN~\cite{schubert2017dbscan} to group the contextualized embeddings.
While, through constructing the contextualized embeddings, the algorithm is more likely to put papers relevant to the exploration context together, it can produce other clusters less related to the contexts.
Besides, the resulting clusters might still have many papers that making it challenging for users to quickly screen them.
To this end, we harness \LLM to curate the clustering results.
Specifically, we prompt \LLM with two questions for each cluster: ``Is this cluster related to the research topic and the selected question?'' and ``Is there a clear division between papers that a subclustering would be helpful?''
Based on the results, we remove irrelevant clusters and decide whether to divide any existing clusters further.
We still keep the papers under the removed clusters in a ``possibly relevant paper'' list in \system for participants to browse if they want.
This design decision is driven by the observation that researchers would not only search on purpose but also be interested in serendipitous information discovery~\cite{foster2004nonlinear}.

\textbf{Supported Interactions.} 
Upon selecting an EQ for which papers have already been queried, \system presents the extracted research themes. 
It displays the disciplinary distribution and key phrases for papers within each theme (\autoref{fig:system}-E) and links the key phrases to their respective wiki pages via DBPedia Spotlight~\cite{mendes2011dbpedia} when possible. This feature aids users in demystifying unfamiliar terms. Users can delve into themes to review the papers classified under them.
\\

\subsubsection{\textbf{[DG4] Support Sifting Papers}}
\system presents retrieved papers with their disciplines, titles, and meta information such as authors, published year, venue, abstract, and references as well as citations (\autoref{fig:system}-F1).
Users can also click a link button to navigate to the \search page of the papers for further investigation.

Besides, \system highlights the relevant phrases in the title and abstract that are aligned with the current exploration contexts, i.e.,  research topic as well as the selected EQ, as shown in \autoref{fig:system}-F1.
The method of extracting relevant keyphrases is the same as the one in the clustering process described in \autoref{sec:organize}. 
Additionally, to prevent information overload in the user interface, only the sentence from the abstract deemed most relevant to the exploration context is displayed. 
Formally, given an abstract represented as a document consisting of a set of sentences \(D = \{s_1, \ldots, s_n\}\) and the set of concepts in the exploration contexts \(C = \{c_1, \ldots, c_m\}\), the key sentence \(s^*\) is identified by:
\(s^* = \arg\max_{s_i \in D} \left|\left\{c_j \in C | \text{ exists a phrase relevant to } c_j \text{ in } s_i\right\}\right|\).
Users can also expand the key sentence to see the full abstract at will.

\subsubsection{Other Features}
Besides the aforementioned four features that directly target the four design goals, \system also includes some other features to facilitate the \task process.

\textbf{Support Organization of Useful Papers.}
\system allows users to drag and drop themes or papers from the \textit{exploration view} to the \textit{collection view}.
When dropping themes, a collection is created in \system with the same title as the themes and containing the same group of papers.
When papers are dropped, they would be added to collections based on their sources: from a theme or from citations and references of other papers.
Users can edit the title of the collections, move papers from one collection to another also through dragging and dropping, and delete collections.
This organization feature, together with the themes extraction feature, supports the mixed-initiative workflow proposed by \citet{kang2023synergi}.

\textbf{Ranking Citations and Reference.}
\system displays the citations and references of a paper upon users' requests, and these citations and references are organized based on their disciplines (\autoref{fig:system}-F2).
In particular, we ranked the disciplines to prioritize the ones that contain relevant information to the exploration contexts but are less explored according to users' interaction history with the system.
This ranking aims to balance exploration and exploitation~\cite{athukorala2016beyond} and support \textbf{DG1}.
We define users' engagement level in each discipline as \(U_d = p_d + q_d\), where \(p_d\) is the number of papers collected and \(q_d\) is the number of EQs queried in discipline \(d\). 
The exploration score, \(E_d\), is set to \(1 / (U_d + 1)\) to encourage exploring lesser-engaged disciplines.
The relevance of each discipline, \(V_d\), is computed as the average cosine similarity between the paper embeddings in \(d\) and the user-specified research topic embedding. 
The combined score \(C_d\) for ranking disciplines is simply computed as \(C_d = \beta V_d + E_d\), where \(\beta\) is set to 1 in our prototype.
For papers under each discipline, we rank them by their cosine similarity score with the users' topic of interest.

\section{Study 1: Usability Evaluation}
To evaluate \system, we first conducted a usability study on \system with a focus on how useful and easy it is to use \system in helping researchers search papers regarding an interdisciplinary research topic.
Specifically, we explore the following research questions:
\begin{enumerate}
    \item[\textbf{RQ1}] Can \system help researchers gather more comprehensive literature? 
    \item[\textbf{RQ2}] Does \system improve user efficiency in the \task process?
    \item[\textbf{RQ3}] How do users perceive the usefulness and experience of interacting with \system??    
\end{enumerate}


\subsection{Participants}
We recruited 12 participants (six females, five males, and one prefer not to say) from social media. 
They are all postgraduate research students.
Eight participants have 1-3 years of research experience, three have 3-5 years, and one have 5-10 years.
Their research directions include HCI (6), visualization (2), machine learning (2), programming language (1) and cognitive science (1). 
All but one participant confirmed that they have experience in interdisciplinary research.
Additionally, five participants suggested they often use AI tools, including Consensus~\footnote{\url{https://consensus.app/}}, Elicit~\footnote{\url{https://elicit.com/}}, Perplexity~\footnote{\url{https://www.perplexity.ai/}} and Paper Digest~\footnote{\url{https://www.paperdigest.org/}}, for searching literature. 

\subsection{Protocol}

We adopted a within-subject experiment design.
Participants were asked to complete \task tasks under two conditions in a counterbalanced order:  the system condition (using \system) and the baseline condition (using tools they usually use).
We prepare two interdisciplinary research topics for participants to perform \task:
\begin{itemize}
\item Develop digital teachers to mentor high school students in rural areas, focusing on training their problem-solving capabilities. We refer to this task as \textbf{AI for education} in later texts.
\item Help designers use AR to creatively document intangible cultural heritage, ensuring its preservation for future generations. We refer to this task as \textbf{AR for ICH} in later texts.
\end{itemize}


Participants were instructed to widely survey papers from various academic fields for each topic.
In both conditions, participants were required to complete an outline in Google Docs as the task outcomes. 
In the outline, participants needed to list the research papers they found relevant and group them into topics.
Participants were also asked to write a short explanation of how the topics related to the given project.

Before the task starts in the system condition, we first provide a 5-minute tutorial on \system and give 5 minutes for participants to try \system.
Then, they are given 30 minutes to search papers and complete the outline for each task.
After each task, they are asked to complete a corresponding questionnaire.
Upon completing both tasks, we ask participants to share their experience in using \system.

\subsection{Measurement}

To answer \textbf{RQ1}, we compare participants' outlines in the two conditions from three aspects:
statistics of the outlines, participants' subjective ratings, and ratings from external experts.
For the statistics of the outlines, we computed the number of topics and papers in each outline.
For the participant's ratings, in the questionnaire after each task, participants self-rated the outline's helpfulness, comprehensiveness, organization, and relevance using the 7-point Likert Scale.
Moreover, before each task started, we asked participants to tell us what they planned to search for during the task.
Subsequently, upon completion of the assigned task, we directed the participants to provide ratings indicating the extent to which they identified relevant literature \textit{they had planned to search}, as well as the degree to which they discovered relevant literature that \textit{had not been part of their initial search strategy}.
These ratings help to assess whether \system achieves the goal of expanding researchers' exploration scope.
For the expert ratings, we recruited two experts to score the outlines.
One is a final-year PhD student, and another is an assistant professor.
Both are HCI researchers and have experience in multiple interdisciplinary research projects widely covering AI, XR, education, and design.
They evaluated the participants' outlines while blind to the condition using a 10-point Likert scale by rating the following statements:
\begin{itemize}
    \item Overall outline helpfulness: ``\textit{I found the outline is overall helpful for developing the research project.}''
    \item Outline comprehensiveness: ``\textit{I believe this outline covers a comprehensive set of relevant topics.}''
    \item Topic relevance for each topic: ``\textit{I found this topic relevant to the project.}''
    \item Topic support for each topic: ``\textit{I found the topic to be well-supported by papers under it.}''
\end{itemize}
We refer to \citet{kang2023synergi} in designing these statements except for the outline comprehensiveness.
We add the outline comprehensiveness as a measurement as an important design goal of \system is to assist researchers in gathering highly-scatter knowledge.
Before their evaluation, experts first engaged in a session to discuss how to give specific points.  
Experts' given ratings were averaged as the final score.

To answer \textbf{RQ2}, we measure the task workload using the NASA task load index~\cite{hart1988development}.
We also measure participants' knowledge gain during the search process, as researchers typically need to learn knowledge from other disciplines on the fly in \task.
Participants rated their knowledge on the given research topic using a seven-point Likert scale before and after the task, and we take the change in self-rated knowledge as the knowledge gain~\cite{palani2022interweave}.
They also rate their self-efficiency during the task.

To answer \textbf{RQ3}, we adopt the technology acceptance model ~\cite{venkatesh2008technology} to assess the perceived usefulness, ease to use, ease to learn, workflow compatibility of \system, and participants' intention to use it.
Participants also rate their perceived helpfulness of \system to key stages of \task, including orienting to various academic fields, breadth searching papers, filtering irrelevant papers, and making sense of the connection between the papers and the focused research project.


\subsection{Findings}
For statistical results presented in this section, we used the paired samples t-test for comparison between conditions, unless the Shapiro-Wilk test of normality indicated a deviation from normality. 
In cases of non-normal data, we utilized the Wilcoxon Signed Rank test.

\begin{figure}
    \centering
    \includegraphics[width=1\linewidth]{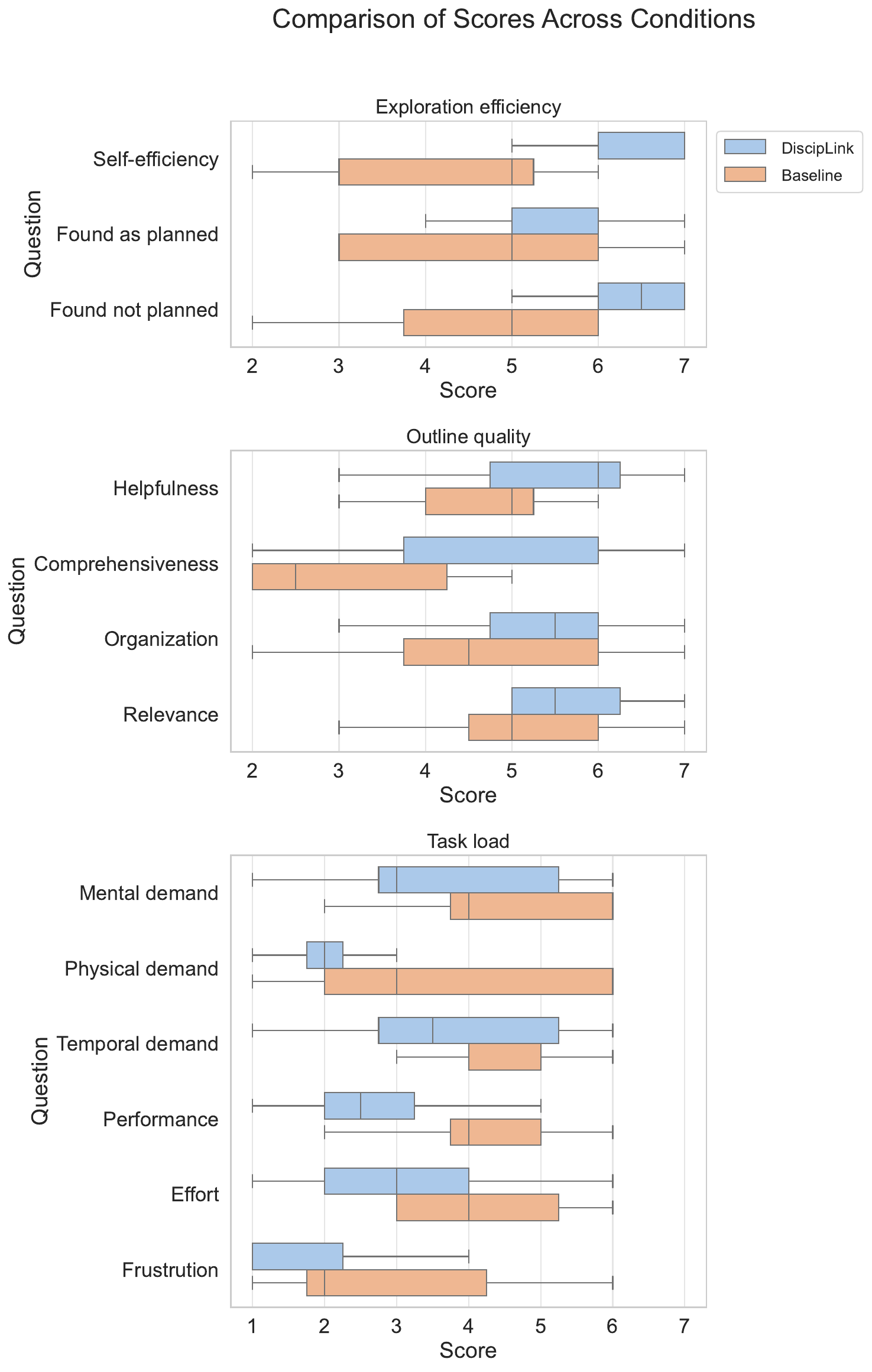}
    \caption{Comparison of participants' self-ratings between the two conditions. In particular, 1 point of ``Performance'' means success, and 7 means failure.}
    \label{fig:user-score}
\end{figure}

\subsubsection{[RQ1] Quality of the outline}
Participants in the system condition, on average, collected 5.08 topics (SD=2.28), more than the baseline condition (M=3.92, SD=1.51), but there are no significant differences (\(p\)=0.057).
There are also no significant differences regarding the number of papers (the system condition: M=20.58, SD=14.39; the baseline condition: M=17.83, SD=10.31; \(p\)=0.58).

The experts ratings indicated significantly higher outline helpfulness of the system condition (M=6.67, SD=1.09) than the baseline condition (M=6.04, SD=0.96, \(p\)=0.003).
Similarly, experts' ratings of outline comprehensiveness are significantly higher in the system condition (M=6.63, SD=1.23) than in the baseline condition (M=5.83, SD=0.69, \(p\)=0.02).
Nevertheless, there are no significant differences regarding averaged topic relevance and averaged topic support for the two conditions.

For participants' subjective ratings (shown in \autoref{fig:user-score}), there are no significant differences between the self-report outline helpfulness (\(p\)=0.16), organization (\(p\)=0.31), and relevance (\(p\)=0.18).
However, participants reported higher comprehensiveness of the outline in the system condition (M=5.00, SD=1.60) compared to the baseline condition (M=3.08, SD=1.31, \(p\)=0.002).
Besides, there is no significant difference between how well participants consider they found literature they plan to search (\(p\)=0.07).
But regarding how well they found relevant literature they originally did not plan to search, the system condition (M=6.08, SD=1.44) is significantly better than the baseline condition (M=4.67, SD=1.44, \(p\)=0.05).

These findings suggest that \system is beneficial in helping researchers to gather comprehensive literature from various disciplines as intended.

\subsubsection{[RQ2] Search process}

Participants in the system condition reported significantly larger learning gains (M=2.50, SD=2.468) compared to the baseline condition (M=1.08, SD=1.56, \(p\)=0.02), suggesting \system helps users learn more during the exploration.
Also, they self-reported higher efficiency in the system condition (M=6.18, SD=0.94) compared to the baseline condition (M=4.33, SD=1.50, \(p\)=0.01).
Participants exhibited a significantly higher task workload in the baseline conditions based on the averaging score of the NASA TLX items (M=4.01, SD=0.74) compared to the system condition (M=2.93, SD=0.97, \(p\)<0.01).
For the individual items (illustrated in \autoref{fig:user-score}), the system condition is significantly better than the baseline condition in regard to the physical demands (\(p\)=0.01), performance (\(p\)=0.03), effort (\(p\)=0.02), and frustration (\(p\)=0.04).
These results suggest that \system improves participants' search efficiency, lowers their workload, and helps them effectively learn more in the \task process.

\begin{figure}
    \centering
    \includegraphics[width=1\linewidth]{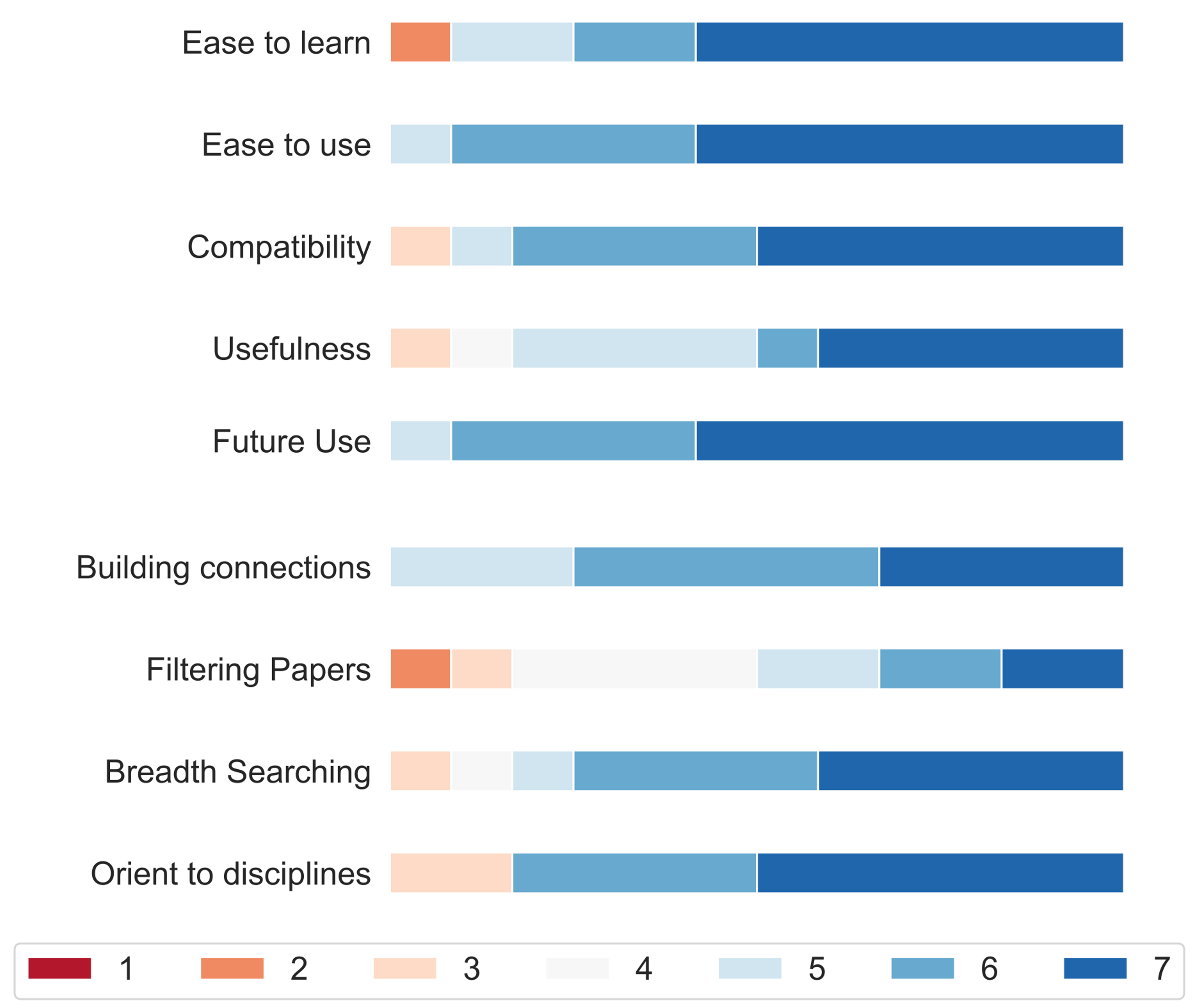}
    \caption{Participants ratings of \system.}
    \label{fig:rating}
\end{figure}

\subsubsection{[RQ3] User experience}
Participants generally held positive attitudes towards the usefulness, ease of use, ease of learning, and workflow compatibility, as shown in \autoref{fig:rating}.
Most participants appreciated the generated EQs.
P12 noted that ``\textit{When I came to this new field, I didn't know where to go and how to search on Google. There might be some terms in this field, and then it [\system] could help me ask some questions. It is difficult for me to ask professional questions in this way.}''

Nevertheless, some participants complained that \system provides too many EQs and thus adds load to users: ``\textit{It has many [EQ] recommendations, and I may still need some efforts to screen them. I need more time to see if this is relevant, and it is unlikely to be completed very well in a relatively short time.}'' (P06).
P07 expects the EQs to be better connected, ``\textit{There may be a connection or overlap between them, but it didn't tell me.}''.

Besides, as illustrated in \autoref{fig:rating}, there were mixed opinions among participants regarding \system's effectiveness in filtering papers. 
Those who viewed \system positively highlighted its ability to deliver relevant themes and papers aligned with the selected EQs. (``\textit{I think the results of the query are very accurate. I can’t get such accurate results using Google Scholar because I feel that it only matches titles. You can search more accurately with this [system].}'', P11) and appraise the keyphrase highlighting (``\textit{The papers provided by other search engines may not be relevant, and I don't know exactly what relevant keyword it has. But you can clearly see the fields, the key phrases [in \system]}'', P09).
However, some participants felt challenged by the volume of papers \system recommended, pointing out the difficulty of effectively screening them within the experiment time, ``\textit{Because it recommends a lot, I still need manual effort to screen them. I can't complete it very well in a relatively short period of time.}'' (P6)

We also examined what tools participants used in the baseline condition.
We found that all participants used Google Scholar, while P6 and P10 used tools including LitMaps and Connected Papers to check citation networks of papers, and P01, P08, P11, P12 attempted AI-powered literature search tools including Consensus, Elicit, Paper Digest, and Perplexity.
Compared to \system, participants suggested the tools displaying citation networks offer better visual salience on key papers but provide limited assistance in guiding search directions.
P10 noted that ``\textit{Some of them [papers] were clustered together [in the citation network], and then I took a quick look at those. [...] But I just picked the standout ones and need to read carefully to proceed, while your tool is much better for getting a comprehensive exploration quickly}.''
Regarding AI-powered search tools, P01, P11, and P12 initially tried them but quickly reverted to Google Scholar due to a lack of helpful answers. Conversely, P08 found Paper Digest useful for creating outlines but felt that ``\textit{Your system has several exploration levels, allowing me to understand the paper I gave better. In the summary generated by Paper Digest, I am more about reading and accepting papers, and my understanding of these papers may not be as clear.}''
This feedback illuminates the unique strengths of \system in unfolding the \task process.

\section{Study 2: Case Study}
Encouraged by the results of study 1, we further conducted an open-ended exploratory study with experienced interdisciplinary researchers to get qualitative feedback on the potential benefits and limitations of \system in \task.

\subsection{Methodology}
\subsubsection{Participants}
We recruited seven researchers (two females and five males) who are focused on interdisciplinary research, including one assistant professor, two postdoctoral researchers, and four senior PhD students.
Participants are from various backgrounds, including AI for healthcare, data storytelling, social psychology, AI for science, music technology, and AI for education.

\subsubsection{Protocal}
We recommended that participants choose projects for which they haven't fully devoted themselves to searching for relevant literature yet, in order to explore those projects in this study.
If they do not have such projects or they feel uncomfortable doing that, participants could also explore papers under their other interdisciplinary projects.
Participants were informed that none of the information on their research projects would be reported.

After getting the consent of participants, at the beginning of the study, we gave a tutorial on \system, as the one in Study 1, and suggested participants get familiar with the system through a 5-minute practice with a given research topic.
We then asked participants to explore their interdisciplinary projects using \system.
We encouraged participants to think aloud and explore \system for at least 20 minutes.
Participants are allowed to explore \system for a longer time if they are willing to do so.
After the open-ended exploration, we conducted semi-structured interviews with participants.
The interview questions cover participants' general experience with \system, whether \system helps in uncovering relevant knowledge in multiple disciplines, whether \system found helpful papers, whether \system helps in screening papers, and how well \system can fit into their daily \task process.
We also asked whether participants felt the level of automation was appropriate.

\subsubsection{Data Collection and Analysis}
All study sessions were audio-recorded with the informants' consent. 
The audio recordings were transcribed using automated tools and subsequently manually verified.
Two members of our research team analyzed the transcripts using thematic analysis~\cite{braun2012thematic}. The coding results were thoroughly discussed and integrated through iterative discussions.

\subsection{Findings}
We report our findings based on the participants' qualitative feedback, focused on those not mentioned by participants in Study 1.
Our reports do not include information about participants' research projects to protect participants' privacy.

\subsubsection{Helpfulness of EQs}
All participants found some EQs generated by \system \textbf{relevant and provided insights} to their research projects, enhancing their literature exploration process.
P7 considered EQs ``\textit{introduces a creative and playful way to pick your sense of curiosity}.''
P3 specifically noted that the EQs aligned closely with the fields most pertinent to his research and matched the disciplines associated with the references cited in his paper.
He noted how exploring these EQs can be helpful to his paper: ``\textit{There are some [EQs] that are relevant and can be helpful for developing the introduction part of the paper. There are also three or four questions that can be explored to develop follow-up future studies or some other discussions.}''

Besides, P4 and P7 both mentioned that the generated EQs would particularly \textbf{be beneficial to researchers in the onboarding stage} of interdisciplinary research.
P4 found the EQs are related to ``\textit{the core knowledge in each field}''; thus ``\textit{it is very helpful for understanding knowledge for first-year Ph.D. students}.''
Similarly, P7 commented that ``\textit{most of the researchers, including myself, have problems with the onboarding process. So, these types of applications or surveys could help us with the onboarding process and also learning the new research area step by step.}''

Moreover, most participants found that \system \textbf{streamlines the \task workflow that aligns with and augments their original workflow}.
P1 noted that:

\begin{quote}
    In the past, I would think of directions in different disciplines and search for them. This process was in my mind, but it would definitely not be so comprehensive. It (\system) would ask some very related questions that I had not thought of before.
\end{quote}

An exception is P6, who usually did not search literature by himself.
He collaborated with experts from other disciplines and read the literature they suggested.
However, he also found \system helpful in his workflow:
\begin{quote}
    Sometimes, you don't want to follow the advice of experts exactly. What experts say is not always correct. [...] This tool is very useful when I want to develop something new [beyond their strengths].
\end{quote}

\subsubsection{Various expectation of EQs}
\label{sec:expectation}
Although participants generally found EQs helpful, they also found some EQs did not meet their expectations.
P4, whose background is chemistry and physics, found that all EQs start with ``How'' or ``What'', but not ``Why''.
He considered ``\textit{asking why}'' is more important as ``\textit{
Understanding the underlying causes is important to conducting my research}''.

Another concern is about \textbf{the connection between the generated EQs and their projects}.
Interestingly, there are diverging opinions from participants on this.
P1, P3, and P6 consider some EQs as not good because the connection between these EQs and their projects is loose and only covers a portion of the key concepts in the projects.
P1 noted that ``\textit{some questions are simply digging into one field, and it seems that they are less relevant to my interdisciplinary project.}''

However, P5 and P7 desired more EQs to delve deeply into other disciplines and not strongly connected to their own projects.
One reason is the lack of existing research on their project; thus, EQs strongly connected to the project ``\textit{won't be able to help me find something very useful.}'' 
Another reason is the need to delve deeply into other disciplines of research, as explained by P5 with her previous experience:

\begin{quote}
    When I was doing [a certain concept] [related interdisciplinary research], I had to read a lot of psychological literature to see how psychologists used [a certain concept]. [...] If I don’t read real psychological papers and do it directly, then I will always read second-hand literature. I'll never know why people chose that model, and it is problematic.
\end{quote}

Moreover, some participants believed \textbf{the provided EQs should be different in different stages of the interdisciplinary research}.
P2 elaborated on the development of the exploration needs and suggested adding more customization over the EQs generation:
\begin{quote}
    When you first explore this [interdisciplinary] field, you can have more possibilities [...] But I'm actually pretty sure I only want papers with [a concept] and [a concept]. I think before [providing EQs], I can have a choice. For example, am I a person who is exploring this [interdisciplinary] topic for the first time, or am I actually an expert in this topic?
\end{quote}



\subsubsection{\system supports human-AI co-exploration}

Participants generally considered the level of automation to be appropriate, and they were able to engage and direct the exploration.
We observed \textbf{how participants actively tailor the EQs for their own goals}.
For example, P1 edited EQs by adding concepts more related to his project to strengthen the connection between the found papers and the project; for P7, the original question was about how an intervention affects an element, and P7 edited the question by adding more elements could be affected to enlarge the search scope.
Participants also initiated the creation of EQs based on their insights during the exploration process.
P1 found ``gamification,'' and P2 found ``music'' as elements that they hadn't considered in their own projects but were relevant through browsing retrieved papers, and they created EQs based on these findings.

Additionally, P7 highlighted how the language framing of the EQs encourages him to tailor the EQs and join the co-exploration:

\begin{quote}
    If your system was giving me very sophisticated questions. Then, it might restrict my own creativity. But because the questions are broad, easy-to-understand. I think it helps me also think by myself. And these [EQs] can act like hints.
\end{quote}

We also found \textbf{most of the participants not only rely on the provided themes and the corresponding papers in \system to seek information, but also deliberately attempt more ways to discover papers}, including digging into the ``possibly relevant paper'' list, the citation and references of papers, and even browse the recommended paper list in the \search paper page after they clicked the link of papers in \system.
This behavior aligns with the nonlinear model that researchers would deliberately increase the breadth of the search in \task~\cite{foster2004nonlinear}.

When asked about their sense of control during the exploration process with \system, \textbf{all participants affirmed they were in control}. 
They attributed this to having a clear objective for their exploration to advance their own projects. 
Thus, their engagement with the system's recommendations was driven not by the system's guidance alone but by the perceived value of these suggestions to their projects. 
P6 encapsulated this sentiment by stating, ``\textit{I have been deeply thinking about this research topic for a while. It's always on my mind. AI cannot diminish my goal or change my focus.}''




\section{Discussion}
This paper explores the potential of LLMs in supporting interdisciplinary information seeking and contributes to the growing body of research on human-LLM interaction across various domains~\cite{zheng2024charting, shi2023nl2color, liu2023selenite, lee2024paperweaver, lam2024concept}. It is worth noting that there are differing opinions in academia regarding the role of LLMs in enhancing search systems.
\citet{metzler2021rethinking} envision that future search systems can build on LLMs to offer answers to user queries instead of references that require users to further explore.
Conversely, \citet{shah2022situating} argue against ``dropping a LLM-based agent as a one-size-fits-all solution'' and emphasize the importance of user interaction with information in search processes, which enables activities such as learning and serendipity.
Our work aligns with \citet{shah2022situating}'s perspective by offering a human-AI co-exploration solution that integrates LLM support while enabling rich user interaction during the search process.
In this section, we discuss the lessons learned in designing and evaluating \system, as well as its limitations and potential directions for future work.

\subsection{Exploratory Questions as Share Representation in AI-powered \task}
In \system, we conceptualize exploratory questions as a shared representation between users and AI to facilitate co-exploration. 
The concept of shared representation, as introduced by \citet{heer2019agency}, refers to the use of a representation, be it textual or visual, in human-AI interaction that is mutually intelligible to both humans and AI. 
This shared foundation allows for collaborative contributions to the task at hand, thereby striking a balance between automation and user agency. 
Participants in Study 1 infrequently modified the EQs, possibly due to time constraints. 
However, during open-ended exploratory studies, we noticed participants customizing the generated EQs and creating their own, showing the potential of EQs to act as a shared representation that encourages engagement from both humans and AI.

Feedback from participants in our first two studies was generally positive regarding the utility of EQs. Yet, some noted the lack of visible connections between EQs, particularly when they span across disciplines and overlap in themes after querying. 
\system currently does not highlight these links by presenting EQs in a list.
To address this, future efforts could consider better supporting the curation and exploration of EQs.
One possible direction is to consider the existing research on curating examples for creativity support, and visually contextualizing the EQs when they are generated~\cite{lupfer2016patterns, webb2016free, chan2024formulating}.

\subsection{Support Learning with Human-AI Co-Exploration}
As noted in the literature~\cite{palmer2010information, foster2004nonlinear} and commented by our participants in study 2 (\autoref{sec:expectation}), learning in \task helps researchers conduct interdisciplinary research over a long period of time.
This insight underpins our decision to create a human-AI co-exploration workflow rather than merely employing AI agents to autonomously compile research papers for scholars~\cite{xiao2023autosurveygpt}.
Our approach is in line with the idea of AI aiding human cognitive processes~\cite{miller2023explainable, gajos2022people}. 
Feedback from study 2 indicates that \system not only piques user curiosity but also facilitates their initiation into interdisciplinary research, thereby supporting learning within \task.

However, much of the existing research on human-AI collaboration has prioritized enhancing collaboration to boost immediate performance outcomes~\cite{ma2023should, bansal2021does, cai2019human, ma2024towards}, often focusing on short-term metrics such as the accuracy of decision-making with AI support. While these immediate results are valuable, we advocate for a broader focus within the field, emphasizing support for enduring outcomes of human-AI collaborations. This includes promoting long-term project success (as in interdisciplinary research), enhancing worker well-being, and contributing to societal benefits.

\subsection{Contextualization in \task}

Interdisciplinary research often involves sifting through a vast array of potentially relevant papers. However, the specific contexts of researchers guide them in concentrating on particular studies. 
\system acknowledges this by using the user-provided research topic and selected EQs as a framework for offering contextualized support: recommending EQs tailored to the research topic and synthesizing themes from the papers based on these EQs.
Feedback from user studies confirms the relevance of EQs to their research topics and the extracted themes to the EQs.

Yet, findings from our Study 2 also indicate that interdisciplinary researchers expect \system to grasp their research contexts in a more nuanced way, beyond just the topics at hand. 
This includes understanding their preferences regarding the framing of questions (e.g., P4 shows a preference for ``why'' questions over ``what'' and ``how''), their desired exploration scope (e.g., some participants favor an in-depth exploration into other disciplines while others prefer a stronger linkage with their own research project), and their current phase in the research process (e.g., P2 suggested questions could initially be broader and become more targeted as the research progresses). 
Although \system allows for customization of EQs, offering more nuanced EQ recommendations could significantly enhance the user experience by saving time and boosting efficiency.
Therefore, future endeavors should delve deeper into the specific factors of the needs of interdisciplinary researchers to offer more finely tuned exploration and research support.


\subsection{Limitation and Future Work}
\subsubsection{Deepening the Support to the Nonlinear Exploration Process}
The design of \system is motivated by the nonlinear model of \task behavior~\cite{foster2004nonlinear}.
\system supports the nonlinear exploration process primarily through generating EQs for orientation, query expansion for opening, and theme extraction for consolidation.
However, support for researchers could be further extended, especially in the consolidation stage and the transitions between different stages.
For example, \system does not provide direct supports for activities in consolidation such as incorporation and verifying, which is challenging due to the diverging tastes and paradigms across disciplines~\cite{brown2015interdisciplinarity}.
Future work can explore methods to help researchers understand how to use the found papers effectively.
One inspiration could come from research on analogical search~\cite{kang2022augmenting}, which supports researchers in transferring solutions from other disciplines. 
Also, improving the interdisciplinary reading experience through features such as guiding questions or in-situ summarization as suggested by ~\citet{august2023paper} can be helpful.

Moreover, participants indicated that \system is limited in assisting researchers in exploring the citation networks of papers of interest, which could facilitate transitioning from consolidation to opening or orientation. Existing work, such as Synergi~\cite{kang2023synergi} and Relatedly~\cite{palani2023relatedly}, can enhance this support. For example, Synergi can aid in theme extraction from citation networks based on text snippets of seed papers~\cite{kang2023synergi}.



\subsubsection{Scale of the user study}
In our usability experiment, the task scenarios are mainly related to HCI research.
We acknowledge that this design may limit the generalizability of our findings to other research contexts. 
Testing \system in a broader range of scenarios quantitatively would enable us to better evaluate its efficacy in supporting \task.
Additionally, both the usability experiment and the open-ended exploratory study are relatively short in comparison to the substantial amount of time typically required for researchers in \task.
Long-term field studies are needed to fully assess \system's support in \task, and benefits as well as limitations of human-AI co-exploration.
For example, some participants in study 2 exhaustively explore all presented papers by the system, which can be explained by their open-minded \task search habits~\cite{foster2004nonlinear}.
It would be interesting to see whether long-term usage of \system would change such habits. 


\section{Conclusion}
In this paper, we present \system, a novel interactive system designed to enhance the process of interdisciplinary information seeking through human-AI co-exploration. \system leverages the capabilities of large language models (LLMs) to assist users in their exploration efforts by generating exploratory questions (EQs) and facilitating automatic query expansion. It enables users to actively participate in the exploration by customizing EQs. Additionally, \system employs contextualized clustering to extract themes from the gathered papers and annotates key phrases within those papers, aiding users in quickly identifying how these papers relate to their research objectives.
Our evaluation of \system includes two user studies. The first study assesses usability, gathering both quantitative and qualitative feedback from 12 researchers, which indicates that \system significantly enhances researchers' ability to conduct thorough paper searches. The second study examines the potential integration of \system into the daily routines of interdisciplinary researchers, with qualitative findings highlighting the value researchers find in co-exploring with \system. However, these studies also highlight certain limitations of \system, pointing towards areas for future investigation and improvement.



\begin{acks}
This work is supported by the seed fund of the Big Data for Bio-Intelligence Laboratory (Z0428) from The Hong Kong University of Science and Technology. We thank the anonymous reviewers for their constructive comments and our user study participants for their efforts and insights. We also thank Yujie Zheng for helping with the literature review and Chengzhong Liu for providing suggestions during the early system design stage.
\end{acks}

\bibliographystyle{ACM-Reference-Format}




\appendix



\section{Example Outputs of Prompts for Exploratory Question (EQ) Generation and Search Query Expansion}
We design \system to address \textit{the terminology gap} by generating simple EQs to ease sensemaking and search queries with specific terminology to ensure search proficiency.
In this appendix, we provide example outputs of the generated EQs and queries.
\subsection{EQ Generation}
\label{sec-eq-gen}
\system uses EQs to help users explore unfamiliar fields by asking LLM to take expert personas and adding a condition in the prompt that the generated EQs need to be easy to understand.
We tested prompts for EQ generation, including the one used in \system, and two ablation prompts: 1) a prompt without persona; 2) a prompt without the simplification condition.
The prompts are presented in \autoref{tab:eq_prompts}.\\
We used these prompts to generate three EQs under three different relevant disciplines for eight research topics (in total 72 EQs). Part of the results are presented in \autoref{tab:generated_eqs}.
We include all the results in the supplementary material.
These results present hints that using persona makes EQs more concrete to domains and adding a simplification requirement enhances the comprehensibility.


 \begin{table*}[h!]
    \caption{Tested prompts for EQ generation.}\label{tab:eq_prompts}
    \begin{tabular}{lp{12cm}}
        \toprule
Condition & Prompt\cr
        \midrule
        w/o persona & 
You are assisting a researcher who is looking for critical research questions studied in a certain field.\newline
The research idea they are interested in is: \{research\_idea\}.\newline
Please provide a list of questions that:\newline
- Are simple, and easy to understand. Are concise and do not exceed 15 words each.\newline
- Relate directly to the research idea.\newline
- Focus on concepts within the field, \{field\}.\newline
Provide \{num\_rq\} questions.\cr
        \hline
        w/o simplification & 
You are an expert in \{field\}. A researcher is looking for critical research questions studied in your field.\newline
The research idea they are interested in is: \{research\_idea\}.\newline
Please provide a list of questions that:\newline
- Relate directly to the research idea.\newline
- Focus on concepts within the field, \{field\}.\newline
The research idea involves multidisciplinary concepts, but you should provide questions that researchers in \{field\} have studied. Do not give questions that you, as a \{field\} expert, cannot answer confidently.\newline
Provide \{num\_rq\} questions.\cr
        \hline
        DiscipLink & \textbf{You are an expert in \{field\}}. A researcher is looking for critical research questions studied in your field.\newline
The research idea they are interested in is: \{research\_idea\}.\newline
Please provide a list of questions that:\newline
- Are \textbf{simple, and easy to understand}. Are concise and do not exceed 15 words each.\newline
- Relate directly to the research idea.\newline
- Focus on concepts within the field, \{field\}.\newline
The research idea involves multidisciplinary concepts, but you should provide questions that researchers in \{field\} have studied. Do not give questions that you, as a \{field\} expert, cannot answer confidently.\newline
Provide \{num\_rq\} questions.\cr
        \bottomrule
    \end{tabular}
\end{table*}

\begin{table*}[h!]
    \centering
    \begin{tabular}{p{3cm}lp{11cm}}
        \toprule
        \textbf{Research Topic \& Field} & \textbf{Prompt} & \textbf{Generated EQs} \cr
        \midrule
        The relationship between mobile phone use and mental well-being [Social Psychology] 
        & w/o persona 
        & - How does mobile phone use impact social interactions and mental well-being? \newline
        - Does excessive phone use contribute to feelings of loneliness or depression? \newline
        - Can mindful usage of mobile phones improve mental health outcomes? \cr 
        \cline{2-3}
        & w/o simpl. 
        & - How does the frequency of mobile phone use impact perceived social support and interpersonal relationships among different age groups? \newline
        - What role does social comparison via social media on mobile phones play in affecting self-esteem and mental well-being? \newline
        - How do patterns of mobile phone use influence feelings of loneliness and social connectedness in various social settings? \cr
        \cline{2-3}
        & DiscipLink 
        & - How does mobile phone use impact social connectedness and isolation? \newline
        - Do smartphone interactions influence users' self-esteem and self-worth? \newline
        - Can excessive mobile phone use contribute to social anxiety or fear of missing out (FOMO)? \cr
        
        \hline
        Promote sustainable travel choices among urban commuters [Transport Economics] 
        & w/o persona 
        &- How does pricing affect urban commuters' choice for sustainable travel methods?\newline
- What role does convenience play in urban commuters' decision to use public transport?\newline
- How do travel time comparisons influence commuters' preference for car vs. public transport?\cr
        \cline{2-3}
        & w/o simpl. 
        & - What are the elasticity effects of pricing policies (e.g., congestion pricing or parking fees) on urban commuters' travel behavior towards more sustainable modes of transportation? \newline
        - How do investments in public transit infrastructure economically impact commuters' propensity to shift from private car usage to public transportation in urban areas? \newline
        - What are the cost-benefit implications of implementing incentives such as subsidies for electric vehicle purchases or biking to work programs on overall urban travel sustainability? \cr
        \cline{2-3}
        & DiscipLink 
        & - How do fare subsidies impact urban commuters' adoption of public transport? \newline
        - What are the economic benefits of reducing car dependency in cities? \newline
        - How do congestion pricing policies influence commuter behavior and mode choice? \cr
        
        \hline
        Facilitate community engagement in local environmental conservation efforts [Public Policy] 
        & w/o persona 
        & - How can public policy increase community participation in environmental conservation?\newline
        - What policies effectively engage youth in local environmental efforts?\newline
        - How do incentives in policies impact community-led conservation activities?\cr
        \cline{2-3}
        & w/o simpl. 
        & - What policy mechanisms are most effective in incentivizing local communities to participate in environmental conservation efforts? \newline
        - How do local governance structures impact the level of community engagement in environmental conservation programs? \newline
        - What role does public policy play in building partnerships between local governments, non-profit organizations, and community groups to support environmental conservation initiatives? \cr
        \cline{2-3}
        & DiscipLink 
        & - How can policy frameworks enhance public participation in local conservation programs? \newline
        - What incentives can encourage local communities to engage in conservation efforts? \newline
        - Which governance structures best support community-led environmental conservation initiatives? \cr
        
        \bottomrule
    \end{tabular}
    \caption{EQs generated by GPT-4 with different prompts. The GPT-4 version we used is GPT-4-0125.}
    \label{tab:generated_eqs}
\end{table*}


\subsection{Search Query Expansion}
\system first generates pseudo-answers (PA) and then uses PA to prompt for queries (\autoref{sec:query-expansion}).
We tested query expansion with PA generation first and without PA.
For each EQ generated in \autoref{sec-eq-gen}, nine search queries were generated.
Part of the results are presented in \autoref{tab:generated_eqs} and all are included in the supplementary material.
We also compared the concreteness of the generated queries of the two query expansion approaches (with and without PA) using the MRC Psycholinguistic Database~\cite{coltheart1981mrc}.
The average concreteness ratings of the queries generated with PA is 319.56 (SD=52.37), and for the queries generated without PA is 292.42 (SD=39.69).
A comparison between the two conditions regarding the generated queries of each input EQ using paired t-test shows the queries generated with PA is significantly more concrete than the ones generated without PA (\(p\)<0.01).
These results provide evidence on the addition of PA helps inject domain-specific terms into the queries.


\begin{table*}[h!]
    \centering
    \begin{tabular}{p{3cm}lp{11cm}}
        \toprule
        \textbf{EQ} & \textbf{Condition} & \textbf{Generated Queries} \cr
        \midrule
        Does heavy social media use affect code-switching behaviors in multilingual speakers?
        & w/o PA 
        & 1. "heavy social media use code-switching multilingual speakers" \newline
          2. "social media impact multilingual code-switching" \newline
          3. "code-switching frequency social media multilingualism" \newline
          4. "effects of social media on bilingual code-switching" \newline
          5. "digital communication code-switching multilingual" \newline
          6. "online interaction code-switching behavior multilingual speakers" \newline
          7. "social media language switching bilingualism" \newline
          8. "social media multilingual communication behavior" \newline
          9. "impact of social networks on code-switching" \cr
        \cline{2-3}
        & w/ PA 
        & 1. "social media influence code-switching multilingual" \newline
          2. "code-switching normalization social media environments" \newline
          3. "multilingualism social media language practices" \newline
          4. "identity negotiation code-switching social media" \newline
          5. "digital communication multilingual audience design" \newline
          6. "social contexts code-switching social media" \newline
          7. "social capital code-switching online communities" \newline
          8. "strategic code-switching social media multilingual" \newline
          9. "building online communities multilingual code-switching" \cr
        \hline
        How effective are price incentives in shifting commuter preferences towards sustainable travel methods?
        & w/o PA 
        & 1. "price incentives commuter preferences sustainable travel" \newline
          2. "financial incentives sustainable transportation" \newline
          3. "commuter behavior price incentives sustainable travel" \newline
          4. "effectiveness of price incentives on sustainable commuting" \newline
          5. "incentives for sustainable travel modal shift" \newline
          6. "price reductions sustainable transportation methods" \newline
          7. "commuter choices financial incentives sustainable travel" \newline
          8. "economic incentives public transportation uptake" \newline
          9. "reward systems for sustainable commuting" \cr
        \cline{2-3}
        & w/ PA 
        & 1. "Price incentives commuter preferences sustainable travel" \newline
          2. "Discounted fares impact sustainable commuting" \newline
          3. "Subsidies cost-benefit analysis commuting" \newline
          4. "Behavioral economics nudges sustainable commuting" \newline
          5. "Price incentives behavioral change transportation" \newline
          6. "Environmental nudges commuting behavior" \newline
          7. "Longitudinal effectiveness price incentives sustainable transport" \newline
          8. "Infrastructure public awareness sustainable commuting" \newline
          9. "Complementary measures price incentives transportation" \cr
        \hline
        What ethical guidelines should govern the use of robots in elderly care?
        & w/o PA 
        & 1. "ethical guidelines robots elderly care" \newline
          2. "robotics ethics senior care" \newline
          3. "AI ethics in elderly care" \newline
          4. "robot caregivers ethical considerations" \newline
          5. "elderly care robotics ethical standards" \newline
          6. "robot use in elder care ethical issues" \newline
          7. "ethics of humanoid robots in eldercare" \newline
          8. "moral guidelines for caregiving robots" \newline
          9. "ethical framework for AI caregivers" \cr
        \cline{2-3}
        & w/ PA 
        & 1. "patient autonomy elderly robot care" \newline
          2. "dignity human-robot interaction elder care" \newline
          3. "ethical design robots elder care" \newline
          4. "privacy elderly care robots" \newline
          5. "data security ethical concerns elder care" \newline
          6. "surveillance data ethics elderly" \newline
          7. "social well-being companion robots elderly" \newline
          8. "emotional support robots mental health elderly" \newline
          9. "human-robot interaction elderly acceptance" \cr
        \bottomrule
    \end{tabular}
    \caption{Queries generated by GPT-4 with or without PA. The GPT-4 version is GPT-4-0125.}
    \label{tab:generated_queries}
\end{table*}



\end{document}